\documentclass[aps,prd,11pt,onecolumn,notitlepage,tightenlines,superscriptaddress,nofootinbib,preprintnumbers,letterpaper]{revtex4-1}
\pdfoutput=1 
\usepackage{amsmath,latexsym,amssymb,graphicx,hyperref,color,enumerate,url,slashed}
\hypersetup{colorlinks,citecolor= nicegreen,linkcolor= nicered}
\usepackage{color}
\definecolor{nicered}{rgb}{0.7,0.1,0.1}
\definecolor{nicegreen}{rgb}{0.1,0.5,0.1}
\begin{document}

\title{Gravitational Waves From SU(N) Glueball Dark Matter}
\author{Amarjit Soni}
\affiliation{Physics Department, Brookhaven National Laboratory, Upton, NY 11973, USA}
\author{Yue Zhang}
\affiliation{Department of Physics and Astronomy, Northwestern University, Evanston, IL 60208, USA}
\affiliation{Walter Burke Institute for Theoretical Physics,
California Institute of Technology, Pasadena, CA 91125, USA}

\begin{abstract}
\noindent
A hidden sector with pure non-abelian gauge symmetry is an elegant and just about the simplest model of dark matter. In this model the dark matter candidate is the lightest bound state made of the confined gauge fields, the dark glueball. In spite of its simplicity, the model has been shown to have several interesting non-standard implications in cosmology. In this work, we explore the gravitational waves from binary boson stars made of self-gravitating dark glueball fields as a natural and important consequence. We derive the dark SU($N$) star mass and radius as functions of the only two fundamental parameters in the model, the glueball mass $m$ and the number of colors $N$, and identify the regions that could be probed by the LIGO and future gravitational wave observatories.
\end{abstract}
\preprint{NUHEP-TH/16-05}

\maketitle

Understanding the nature of dark matter is an open question of central importance for particle physics and cosmology. One of the simplest models of dark matter is a hidden sector with a non-abelian gauge symmetry. In the case of pure gauge theory, its intrinsic scale where the gauge coupling goes strong dictates the mass scale of the dark matter candidate --- the lightest dark glueball state. The glueball dark matter scenario has been considered within various contexts~\cite{Carlson:1992fn, Faraggi:2000pv, Juknevich:2009ji, Boddy:2014yra, Boddy:2014qxa, Yamanaka:2014pva, Buen-Abad:2015ova, Soni:2016gzf, Forestell:2016qhc, Halverson:2016nfq}. 
As emphasized in~\cite{Soni:2016gzf}, a hidden sector with pure SU($N$) gauge group without any fermions or any other intricacy is motivated by its elegance and simplicity. In the CP conserving case, such a hidden sector contains only two parameters: the intrinsic scale $\Lambda$ (or the lightest scalar glueball mass $m$ which we use more often) and the number of colors $N$. In spite of the very few parameters, we have shown that the model can have a number of non-standard and interesting implications in cosmology~\cite{Soni:2016gzf}. In particular, the dark glueball could be a self-interacting and warm dark matter candidate if $0.01\,{\rm keV}<m<10\,$keV and $10^6>N>10^3$. In this case, the self-gravitation of the dark glueball field is allowed to form boson stars that are much more massive than the sun, $\sim 10^{6}-10^8 M_{\odot}$. We will investigate the consequence of such a possibility in this work. 

The recent observation of gravitational waves at the LIGO experiment~\cite{Abbott:2016blz, Abbott:2016nmj} opens up a new avenue to explore a series of possible astrophysical sources. 
A scalar dark matter field could self-gravitate and form massive and compact objects often called the dark stars.
It is natural to imagine a picture where two dark stars are orbiting each other, radiating gravitational waves, until their distance is reduced to the sum of the two radii when they merge with each other~\cite{Giudice:2016zpa}. 
As the distance gets smaller, the gravitational wave frequency gets higher.
For binary dark stars with equal mass $M$ and radius $R$, there is a highest frequency of the gravitational wave radiation,
\begin{equation}\label{binary}
f_{\rm max} = \frac{1}{2\pi} \sqrt{\frac{G M}{R^3}} \ .
\end{equation} 
If $f_{\rm max}$ frequency lies within the range 50--100\,Hz, the gravitational waves from the dark stars are naturally the target of search at LIGO. Future gravitational wave observatories are expected to be sensitive to lower frequencies. Interestingly, the mass and radius of the dark star, and in turn $f_{\rm max}$, are closely related to the effective potential of the scalar dark matter field.

In this note, we explore a natural consequence of glueball dark matter from SU($N$) gauge theory: dark SU($N$) stars (DSS), {\it i.e.}, self-gravitating and compact configurations of the lightest scalar glueball field. It was shown that with non-abelian gauge interactions the ``geon-like'' configuration does not form by itself~\cite{Coleman:1977yb} but with gravity included a glueball star becomes possible.
Our goal is to derive the properties of DSS based on two models of scalar glueball potentials, and assuming binary DSS systems exist close by enough, identify the region of parameter space which LIGO and future gravitational wave observatories are sensitive to. There has been a long history in the literature of finding the boson star solutions for massive scalar fields without and with interactions~\cite{Feinblum:1968nwc, Kaup:1968zz, Ruffini:1969qy, Mielke:1980sa, Colpi:1986ye, Seidel:1991zh, Jetzer:1991jr, Lee:1991ax, Mielke:1997re, Eby:2015hsq}. It was observed that a repulsive self interaction allows the boson star to become 
even much more massive than the sun~\cite{Eby:2015hsq, Soni:2016gzf}. It was also realized that although for a complex scalar field the boson star solution is static, for a real scalar field the solution oscillates as a function of time~\cite{Seidel:1991zh}. The DSS we consider belongs to the latter case. For the non-interacting real scalar case, it has been solved numerically. In general, the scalar glueball field $\phi(r,t)$ can be expanded in Fourier series of $\cos (2j+1)\omega t$, where $j\geq0$ are integers. The numerical results in Refs.~\cite{Brito:2015yfh, Seidel:1991zh} showed that the series converges very quickly, and is dominated by $j=0$. For simplicity, hereafter, we will make the assumption that the oscillation of the glueball field is dominated by a single frequency
of order the scalar mass, $\omega \sim m$, which is also the case in light of the Bose-Einstein condensation. In this case, the space and time dependence in the glueball field $\phi$ can be factorized. We look for spherical symmetric boson star solutions,
\begin{equation}
\phi(r,t) \simeq \Phi(r) \cos\omega t \ .
\end{equation} 
An oscillating DSS by itself does not radiate gravitational waves, but a binary system made of a pair of DSS will do, especially right before they finally merge into each other.
The gravitational wave frequency that LIGO can measure is $f\sim 50-1000\,{\rm Hz}$.
It is important to note that for the glueball to be a cold enough dark matter candidate, there is a huge hierarchy between $f$ and $\omega$, $f\ll\omega$.
This implies that for a binary system, the time dependence information of individual boson star oscillations will not be measured, but only the time averaged effects instead. This suggests to us to average out the oscillation effects with frequency $\omega$ throughout our calculation, in both the Klein-Gordon equation for $\phi$ and the energy-momentum tensor in the Einstein equation. 

With the time dependence of $\phi$ averaged out (see below), it is sufficient to just consider the time independent part of the metric,
\begin{equation}
ds^2 = - B(r) dt^2 + A(r) dr^2 + r^2 d\theta^2 + r^2\sin^2\theta d\varphi^2 \ .
\end{equation}
For the general Lagrangian for $\phi$,
\begin{equation}
\mathcal{L} = - \frac{1}{2} g^{\mu\nu} \partial_\mu \phi \partial_\nu \phi - V(\phi) \ ,
\end{equation}
the coupled classical Einstein-Klein-Gordon equations (after the time average) take the form
\begin{eqnarray}
&&\frac{A'}{r A^2} + \frac{1}{r^2}\left( 1 - \frac{1}{A} \right) = 4 \pi G \left[ \frac{\omega^2 \Phi(r)^2}{B} \left\langle\sin^2\omega t \rule{0mm}{4mm}\right\rangle + \frac{\Phi'(r)^2}{A}\left\langle\cos^2\omega t \rule{0mm}{4mm}\right\rangle + 2 \left\langle V(\Phi \cos\omega t) \rule{0mm}{4mm}\right\rangle \right] \ , \label{e1} \\
&&\frac{B'}{r A B} - \frac{1}{r^2}\left( 1 - \frac{1}{A} \right) = 4 \pi G \left[ - \frac{\omega^2 \Phi(r)^2}{B} \left\langle\sin^2\omega t \rule{0mm}{4mm}\right\rangle + \frac{\Phi'(r)^2}{A}\left\langle\cos^2\omega t \rule{0mm}{4mm}\right\rangle - 2 \left\langle V(\Phi \cos\omega t)\rule{0mm}{4mm}\right\rangle \right] \ , \label{e2}\\
&&\Phi''(r) + \left( \frac{2}{r} + \frac{B'}{2B} - \frac{A'}{2A} \right) \Phi'(r) + A \left[ \frac{\omega^2\Phi(r)}{B} - \left\langle \frac{1}{\cos\omega t}\frac{d V(\phi)}{d\phi}\right\rangle \right] =0 \ , \label{kg}
\end{eqnarray}
where $'$ ( $\dot{ }$ ) means derivative with respect to $r$ ($t$). The symbol $\langle{\ } \rangle$ means taking the time averaging over the $\phi$ oscillation period, $2\pi/\omega$.

In the next two sections, we will solve these equations for the DSS properties based on two assumptions of the scalar glueball effective potential.

\subsection{\large A. \ $\phi^4$ Potential}\label{secA}

As the first case, we assume a $\lambda \phi^4$ potential for the dark glueball. This is a simplified but interesting case where the same quartic coupling controls both glueball dark matter self-interaction strength and the properties of the DSS. 
We assume $\lambda>0$ so the glueball self-interaction is repulsive, or equivalently, its contribution to the scalar potential energy is positive. Stable boson star configuration could be obtained when this repulsive interaction balances the attraction from gravity. The potential takes the form
\begin{equation}\label{V1}
V(\phi) = \frac{1}{2} m^2 \phi^2 + \frac{1}{4} \lambda \phi^4 \ .
\end{equation} 
In this case, the above Eqs. (\ref{e1}, \ref{e2}, \ref{kg}) can be simplified to
\begin{eqnarray}
&&\mathcal{M}'(x) = x^2 \left[ \frac{1}{4} \left( \frac{\Omega^2}{B(x)} + 1 \right) \sigma(x)^2 + \frac{3}{32} \Lambda \sigma(x)^4 + \frac{\sigma'(x)^2}{4A(x)} \right] \ , \\
&&\frac{B'(x)}{x A(x) B(x)} - \frac{1}{x^2}\left( 1 - \frac{1}{A(x)} \right) = \frac{1}{2} \left( \frac{\Omega^2}{B(x)} - 1 \right) \sigma(x)^2 - \frac{3}{32} \Lambda \sigma(x)^4 + \frac{\sigma'(x)^2}{4A(x)} \ , \\
&&\sigma''(x) + \left( \frac{2}{x} + \frac{B'(x)}{2B(x)} - \frac{A'(x)}{2A(x)} \right) \sigma'(x) + A(x) \left[ \left(\frac{\Omega^2}{B(x)} - 1 \right) \sigma(x) - \frac{1}{2} \Lambda \sigma^3(x) \right] =0 \label{kg2} \ ,
\end{eqnarray}
where we have defined $x= m r$, $\sigma = \sqrt{4\pi G} \Phi$, $\Omega = \omega/m$, $\Lambda = \lambda/(4\pi G m^2)$, and
\begin{equation}
A = \left( 1- \frac{2 \mathcal{M}}{x} \right)^{-1} \ . 
\end{equation} 
From the definition of Schwarzschild metric, $M_{\rm pl}^2\mathcal{M}(x)/m$ is the mass of the star within a radius $x/m$. Note that $\mathcal{M}$ and $x$ are both dimensionless.

\begin{figure}[t]
\centerline{\includegraphics[width=8cm]{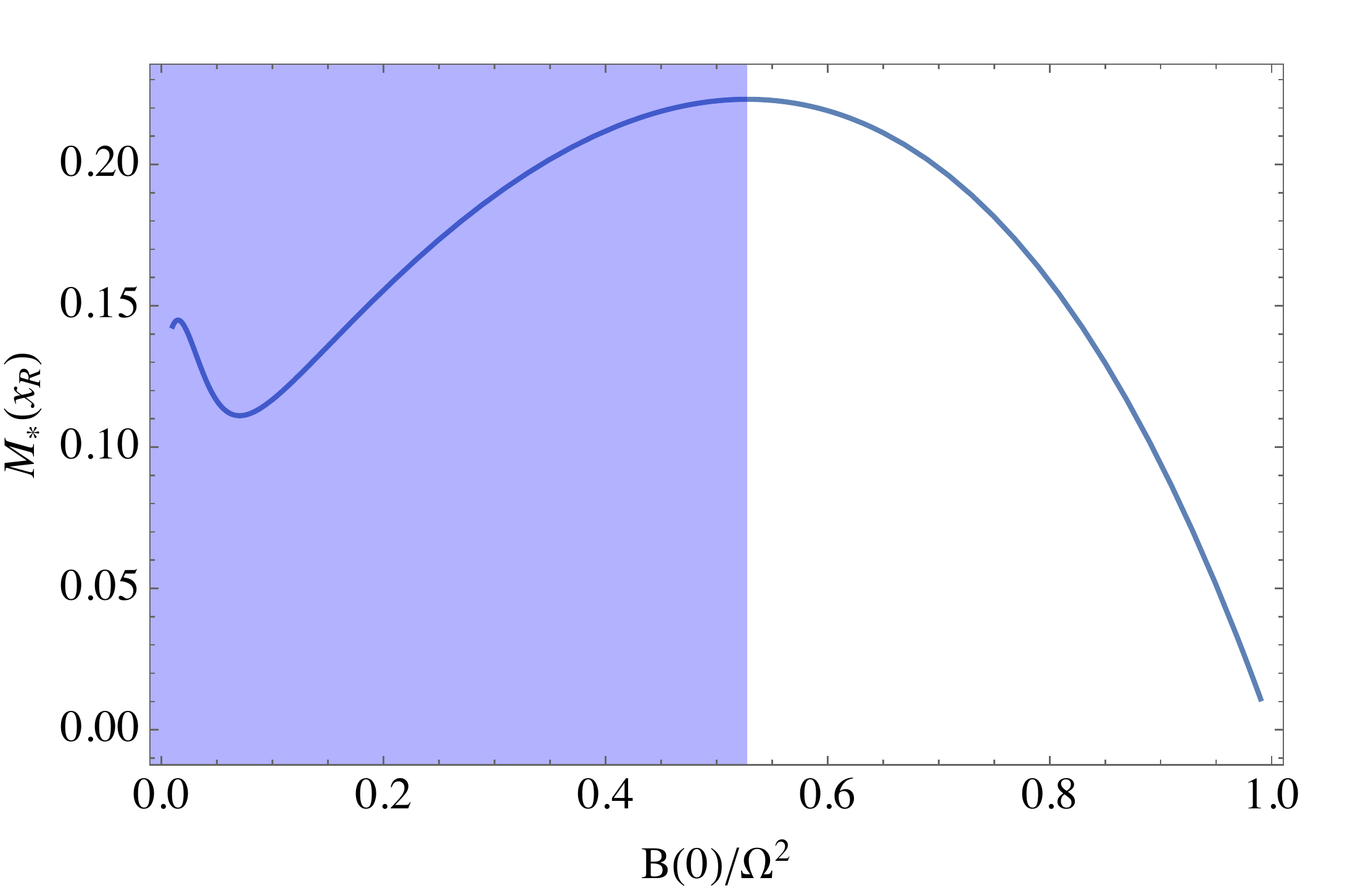}} 
\centerline{\includegraphics[width=8cm]{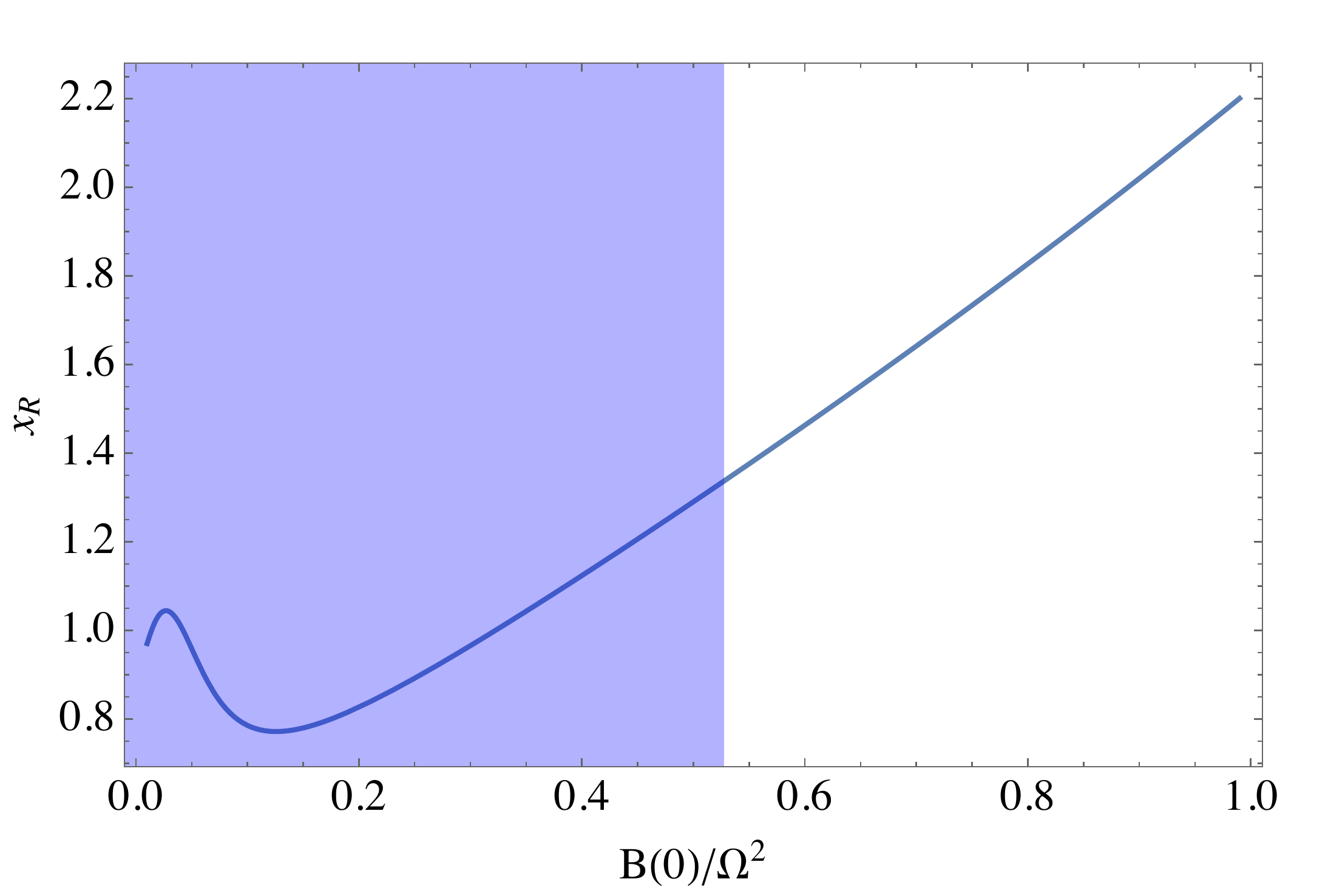}}
\centerline{\includegraphics[width=8cm]{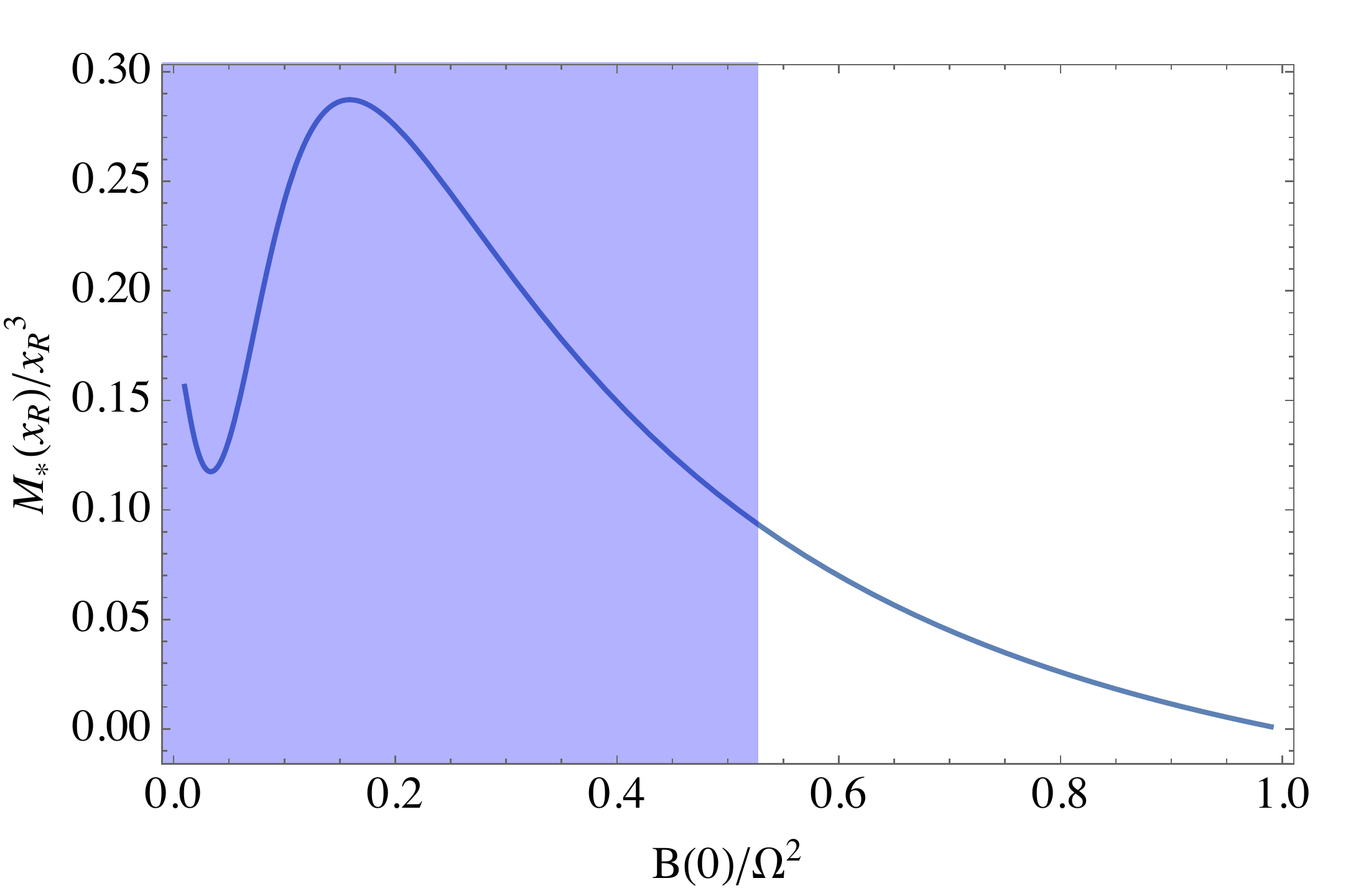}}
\caption{Properties of the DSS by numerically solving the coupled classical Einstein-Klein-Gordon equations with a $\phi^4$ scalar glueball potential, as described in Section A. The upper plot shows the quantity $\mathcal{M}_*(x_R)$, the DSS mass in unit of $\sqrt{\lambda/4\pi}\left(M_{\rm pl}^3/m^2 \right)$, as a function of the boundary condition $B(0)/\Omega^2$. The middle plot shows the quantity $x_R$, the DSS radius in unit of $\sqrt{\lambda/4\pi}\left(M_{\rm pl}/m^2 \right)$, as a function of $B(0)/\Omega^2$.
The lower plot shows the ratio $\mathcal{M}_*(x_R)/x_R^3$ as a function of $B(0)/\Omega^2$.
The blue shaded region is not accessible through the accretion process.}\label{phi4}
\end{figure}

We notice that for glueball dark matter, the model parameters satisfy the condition $\Lambda \gg 1$. In this case, the above equation can be further simplified. Following~\cite{Colpi:1986ye}, we further define
$\sigma_* = \sqrt{\Lambda} \sigma$, $x_* = x/\sqrt{\Lambda}$, $\mathcal{M}_* = \mathcal{M}/\sqrt{\Lambda}$. First, the KG equation (\ref{kg2}) becomes
\begin{eqnarray} \label{kg3} 
\Lambda^{-1}\left[\sigma_*''(x_*) + \left( \frac{2}{x_*} + \frac{B'(x_*)}{2B(x)} - \frac{A'(x_*)}{2A(x_*)} \right) \sigma'_*(x_*)\right] + A(x_*) \left[ \left(\frac{\Omega^2}{B(x_*)} - 1 \right) \sigma_*(x_*) - \frac{1}{2} \sigma_*^3(x_*) \right] =0 \ . \nonumber \\
\end{eqnarray}
In the large $\Lambda$ limit, the first term can be dropped, and we obtain, 
\begin{eqnarray}\label{sigma*}
\sigma_*(x_*) \simeq \sqrt{ 2 \left(\frac{\Omega^2}{B(x_*)} - 1 \right) } \label{kg4} \ .
\end{eqnarray}
This approximate relation is valid until $\sigma_*$ approaches 0, where the second term in Eq.~(\ref{kg3}) vanishes and we could no longer neglect the terms with derivative on $\sigma_*$. With Eq.~(\ref{sigma*}), the two Einstein equations take the form,
\begin{eqnarray}
&&\mathcal{M}'_*(x_*) \simeq x_*^2 \left[ \frac{1}{4} \left( \frac{\Omega^2}{B(x)} + 1 \right) \sigma_*(x_*)^2 + \frac{3}{32} \sigma_*(x_*)^4 \right] \ , \label{e3} \\
&&\frac{B'(x_*)}{x_* B(x_*)} \left( 1 - \frac{2 \mathcal{M}_*(x_*)}{x_*} \right) - \frac{2 \mathcal{M}_*(x_*)}{x_*^3} \simeq 
\frac{1}{2} \left( \frac{\Omega^2}{B(x)} - 1 \right)\sigma_*(x_*)^2 - \frac{3}{16} \sigma_*(x_*)^4 \ . \label{e4}
\end{eqnarray}
We solve Eqs.~(\ref{kg4}, \ref{e3}, \ref{e4}) numerically starting with boundary conditions at the origin $\mathcal{M}_*(x_*=0)=0$ up to the point $x_*=x_R$ where $\sigma_*(x_R) \to 0$. We vary $B(0)/\Omega^2 <1$ as a free parameter. The boson star mass and radius are determined by,
\begin{eqnarray}\label{MR1}
M = \sqrt{\frac{\lambda}{4\pi}} \frac{M_{pl}^3}{m^2} \mathcal{M}_*(x_R)\ , \ \ \ \ \  R =  \sqrt{\frac{\lambda}{4\pi}} \frac{M_{pl}}{m^2} x_R \ .
\end{eqnarray}
Note that from our definition above $x_R$ and $\mathcal{M}_*$ are dimensionless quantities, plotted in Fig.~\ref{phi4}.
 
The results are shown as a function of $B(0)/\Omega^2$ in Fig.~\ref{phi4}. The shaded regions cannot be reached. This can be understood from the picture where the DSS accretes its mass by capturing more and more dark matter particles around it. The mass growth begins from the rightmost toward the left along the curve until it reaches the maximum at $B(0)/\Omega^2=0.53$, with $\mathcal{M}_*(x_R)=0.22$ and $x_R=1.35$.
Thus the mass and radius of the DSS are,
\begin{eqnarray}
M= \sqrt{\lambda} \left( \frac{0.3\,\rm GeV}{m} \right)^2 M_{\odot} \ , \hspace{0.5cm} R = \sqrt{\lambda} \left( \frac{0.3\,\rm GeV}{m} \right)^2 \times 10 \,{\rm km} \ .
\end{eqnarray} 
where $M_{\odot} =2\times10^{30}\,$kg is the solar mass. This corresponds to the highest ratio, ${\rm Max}[{\mathcal{M}_*(x_R)}/{x_R^3}] \simeq 0.09$.
From Eqs.~(\ref{binary}) and (\ref{MR1}), we derive the highest gravitational wave frequency,
\begin{eqnarray}\label{fmax1}
f_{\rm max} = \frac{m^2}{2\pi M_{\rm pl}} \sqrt{\frac{4\pi}{\lambda} {\rm Max}\left[\frac{\mathcal{M}_*(x_R)}{x_R^3}\right]} \simeq 
50\, {\rm Hz} \times \sqrt{\frac{1}{\lambda}} \times \left( \frac{m}{0.05\,\rm GeV} \right)^2\ .
\end{eqnarray}

\subsection{\large B. \ Glueball Potential From Large $N$ Limit}

In general, the scalar glueball potential not only contains the quartic term but also the cubic and higher dimensional interaction terms. In the large $N$ limit, they follow the power counting $\lambda_3 \sim 1/N$, $\lambda_4 \sim 1/N^2$, $\lambda_5 \sim 1/N^3$ and so on.
In this section, we consider a more realistic dark glueball potential based on the large $N$ power counting~\cite{Soni:2016gzf, Forestell:2016qhc}, 
\begin{equation}\label{VlargeN}
V(\phi) = \frac{a_2}{2!} m^2 \phi^2 + \frac{a_3}{3!} \left( \frac{4\pi}{N} \right) m \phi^3 + \frac{a_4}{4!} \left( \frac{4\pi}{N} \right)^2 \phi^4 + \frac{a_5}{5!} \left( \frac{4\pi}{N} \right)^3 \frac{\phi^5}{m} + \cdots \ .
\end{equation} 
The coefficients $a_i$ are order 1 parameters and they could in principle be reliably determined from lattice calculations. To proceed, we will assume that all $a_i=1$. In this case the potential takes the more compact form
\begin{equation}
V(\phi) = \frac{m^4 N^2}{16\pi^2} \left( e^{\frac{4\pi \phi}{Nm}} - \frac{4\pi \phi}{Nm} - 1 \right) \ .
\end{equation} 
We repeat the derivations similar to the $\phi^4$ potential case, and reach the following coupled equations, in analogy to Eqs.~(\ref{kg4}, \ref{e3}, \ref{e4}),
\begin{eqnarray}
&&\frac{\Omega^2}{B(x_*)} \simeq { }_pF_q\left[ \left\{1/2 \right\}, \left\{ 1, 3/2 \right\}, 4\pi \sigma_*(x_*)^2 \rule{0mm}{4mm}\right] \ , \label{20} \\
&&\mathcal{M}'_*(x_*) \simeq x_*^2 \left[ \frac{\Omega^2}{4B(x)} \sigma_*(x_*)^2 + \frac{1}{16\pi^2} \left( I_0 \left(4\pi \sigma_*(x_*)\rule{0mm}{3.5mm}\right) -1 \rule{0mm}{4mm}\right) \right] \ , \label{21} \\
&&\frac{B'(x_*)}{x_* B(x_*)} \left( 1 - \frac{2 \mathcal{M}_*(x_*)}{x_*} \right) - \frac{2 \mathcal{M}_*(x_*)}{x_*^3} \simeq \frac{\Omega^2}{2B(x)}\sigma_*(x_*)^2 - 
\frac{1}{8\pi^2} \left( {\rm I}_0 \left(4\pi \sigma_*(x_*)\rule{0mm}{3.5mm}\right) -1 \rule{0mm}{4mm}\right) \ ,
\end{eqnarray} 
where the field and parameter redefinitions are similar to above, except that here $\Lambda = 1/ (4\pi G m^2 N^2)$. Throughout the parameter space of interest to this study, $\Lambda \gg 1$. The function $I_0$ is the modified Bessel function, and ${ }_pF_q$ is the generalized hypergeometric function.

For deriving Eq.~(\ref{20}, \ref{21}), we have used the relations
\begin{eqnarray}
\left\langle \frac{e^{\frac{4\pi \Phi \cos\omega t}{Nm}} -1}{\cos\omega t} \right\rangle = \frac{4\pi \Phi}{N m} { }_pF_q\left[ \left\{1/2 \right\}, \left\{ 1, 3/2 \right\}, \frac{4\pi^2 \Phi^2}{N^2 m^2} \rule{0mm}{4mm}\right] \ , \hspace{0.5cm}
\left\langle e^{\frac{4\pi \Phi \cos\omega t}{Nm}}\right\rangle = I_0 \left( \frac{4\pi \Phi}{Nm} \right) \ .
\end{eqnarray}

\begin{figure}[t]
\centerline{\includegraphics[width=8cm]{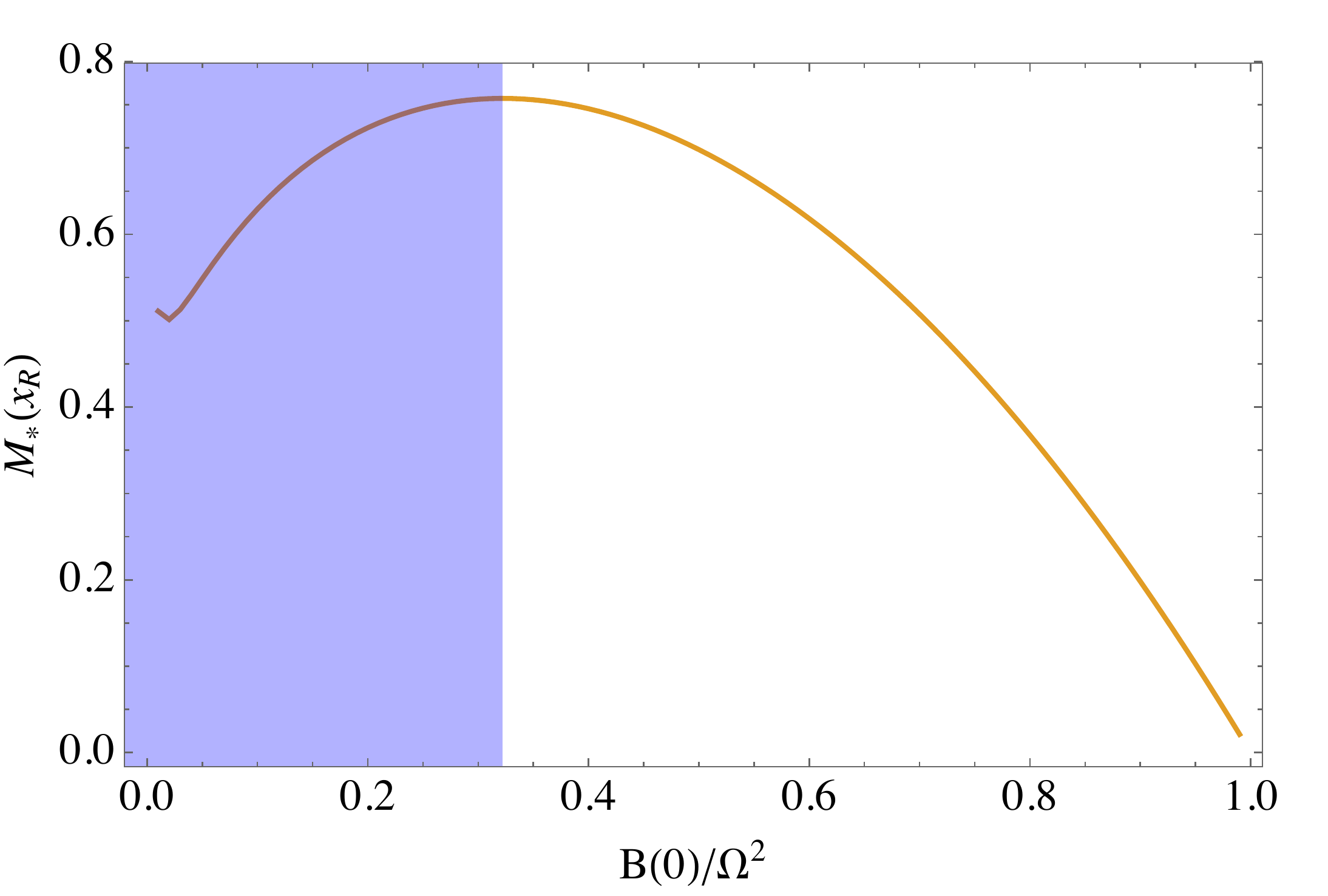}} 
\centerline{\includegraphics[width=8cm]{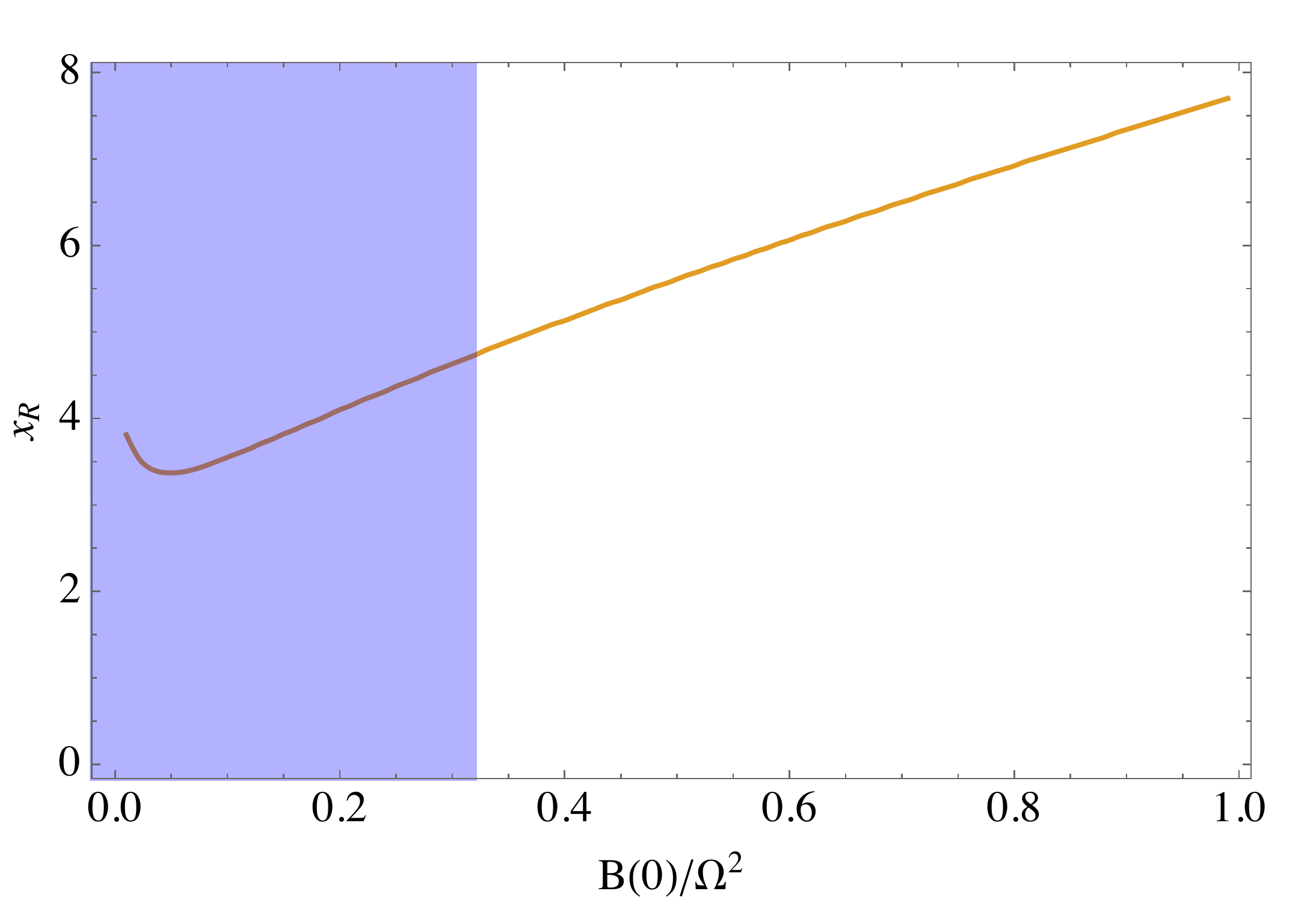}}
\centerline{\includegraphics[width=8cm]{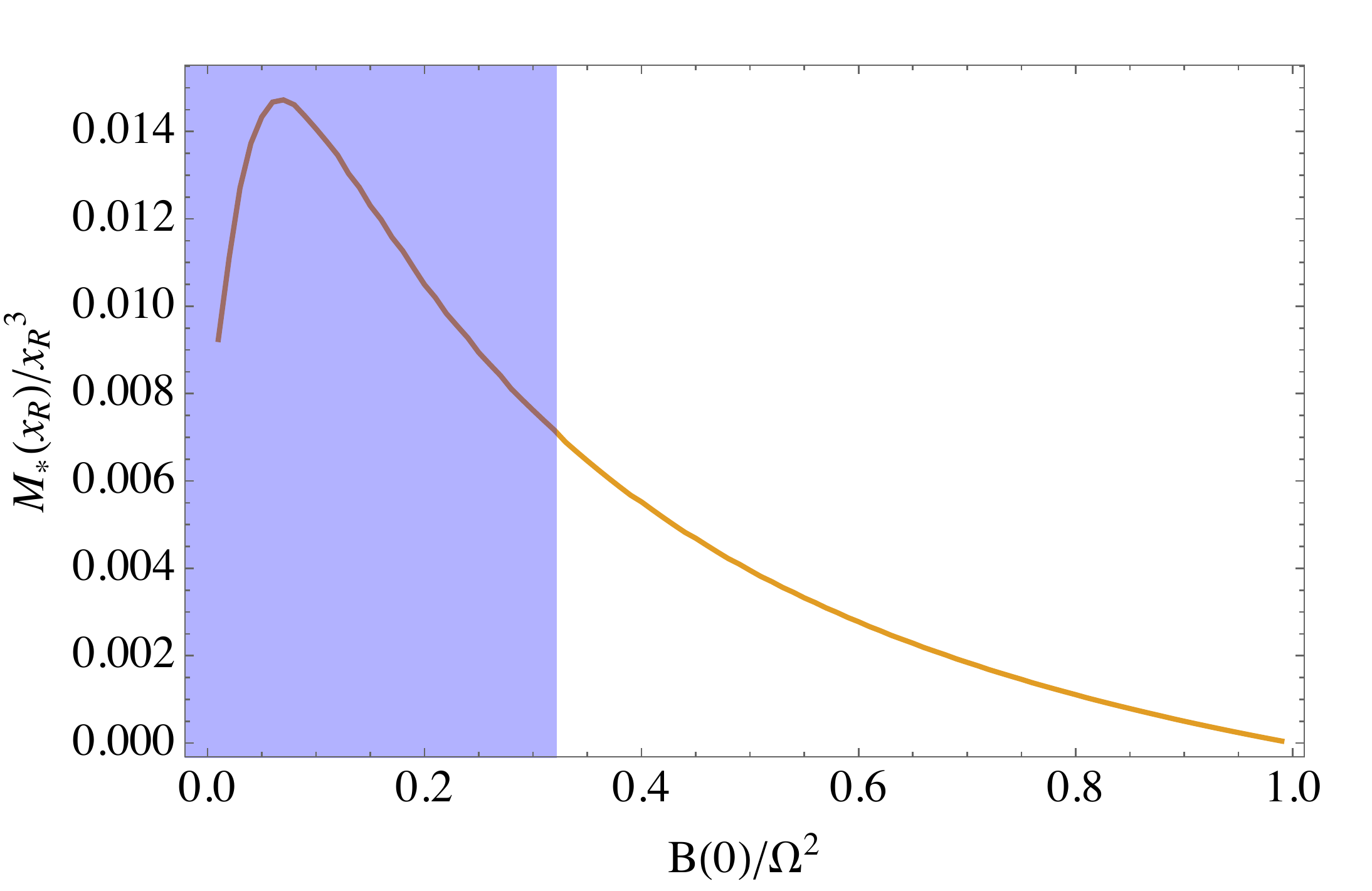}}
\caption{Properties of the DSS by numerically solving the coupled classical Einstein-Klein-Gordon equations with a scalar glueball potential following from large $N$ counting, as described in Section B. The upper plot shows the quantity $\mathcal{M}_*(x_R)$, the DSS mass in unit of $M_{\rm pl}^3/\left(\sqrt{4\pi}  N m^2 \right)$, as a function of the boundary condition $B(0)/\Omega^2$. The middle plot shows the quantity $x_R$, the DSS radius in unit of $M_{\rm pl}/\left(\sqrt{4\pi}  N m^2 \right)$, as a function of $B(0)/\Omega^2$.
The lower plot shows the ratio $\mathcal{M}_*(x_R)/x_R^3$ as a function of $B(0)/\Omega^2$.
The blue shaded region is not accessible through the accretion process.}\label{e^phi}
\end{figure}

\begin{figure}[t]
\centerline{\includegraphics[width=12cm]{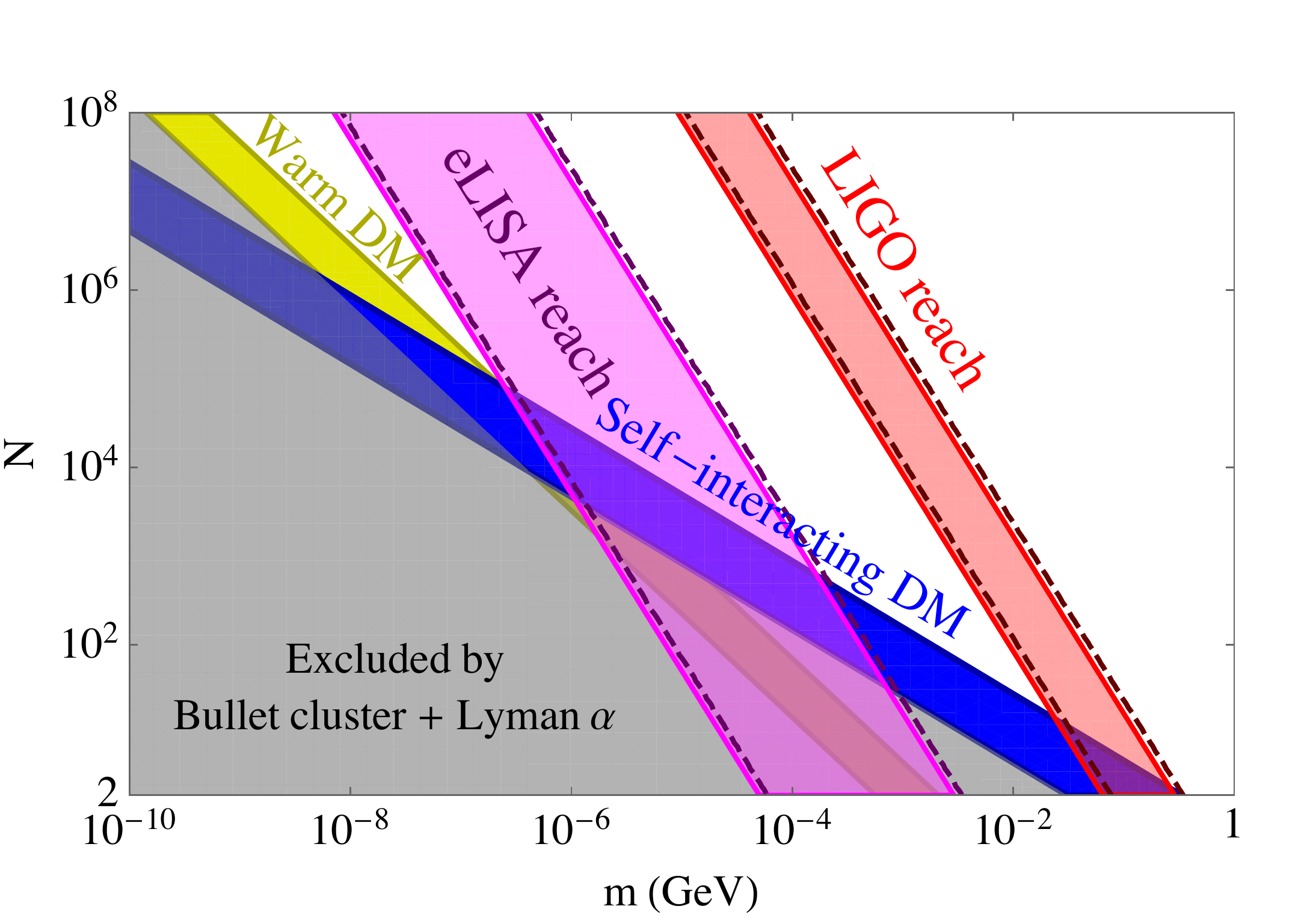}}
\caption{The LIGO experiment could probe the SU($N$) glueball dark matter parameter space, assuming binary DDS exist. 
In the red region, the highest frequency of gravitational wave radiation calculated based on the large $N$ glueball potential in section B, lies between 50--1000\,Hz, thus this region can potentially be within the LIGO sensitivity. 
In the magenta region, the highest gravitational wave frequency from binary DSS is between 0.03\,mHz and 0.1\,Hz and could be probed by the future LISA/eLISA project.
The regions between the dashed lines corresponds the case of $\phi^4$ potential (discussed in section A).
For the same SU($N$) model, in the the yellow band the $3\to2$ annihilation enables the lightest scalar glueball to have the proper free-streaming length to be a warm dark matter candidate, while in the blue band, the $2\to2$ elastic scattering of the glueball dark matter is large enough for it to be a self-interacting dark matter candidate~\cite{Soni:2016gzf}. In the lower left corner, the gray region is already ruled out because of the bullet cluster and Lyman alpha observations.
}\label{LIGO}
\end{figure}

The numerical results are shown in Fig.~\ref{e^phi}. Compared to the $\phi^4$ case, if $\lambda \simeq 1/N^2$, we find that, by varying $B(0)/\Omega^2$, the DSS is allowed to be more massive and at the same time much larger in radius. As a result, the gravitational wave radiated from equal-mass binary DSS has lower frequency.
We find the highest gravitational wave frequency that can be radiated by the binary glueball dark star system corresponds to $B(0)/\Omega^2=0.32$. At this point, the star mass reaches its maximum, with $\mathcal{M}_*(x_R)=0.74$ and $x_R=4.7$. The corresponding mass and radius of DSS are,
\begin{eqnarray}\label{MR2}
\begin{split}
M &= \sqrt{\frac{1}{4\pi}} \frac{M_{pl}^3}{N m^2} \mathcal{M}_*(x_R) = \left( \frac{1}{N} \right) \left( \frac{0.6\,\rm GeV}{m} \right)^2 M_{\odot} \ , \\
R &=  \sqrt{\frac{1}{4\pi}} \frac{M_{pl}}{N m^2} x_R = \left( \frac{1}{N} \right)\left( \frac{0.6\,\rm GeV}{m} \right)^2 \times 10 \,{\rm km} \ .
\end{split}
\end{eqnarray}
Interestingly, if $m\sim 1\,$GeV and $N\sim\mathcal{O}(1)$, the DSS has the typical mass as a massive compact halo object (MACHO).
On the other hand, \cite{Soni:2016gzf} showed that, for the dark glueball to be both self-interacting and warm dark matter candidate, the favored ranges of parameters are $m\sim 0.01-10\,$keV, $N\sim 10^6-10^3$. Following (\ref{MR2}), this corresponds to the highest DSS mass in the range $10^6-10^9M_{\odot}$ and the lowest DSS radius in the range $10^2-10^5R_\odot$, where the solar radius is $R_\odot = 7\times10^5\,$km. 

With ${\mathcal{M}_*(x_R)}/{x_R^3}\simeq0.007$, we find the highest gravitational wave frequency is given by
\begin{eqnarray}\label{fmax2}
f_{\rm max} = \frac{m^2}{2\pi M_{\rm pl}} \sqrt{4\pi N^2 {\rm Max}\left[\frac{\mathcal{M}_*(x_R)}{x_R^3}\right]} 
\simeq  50\, {\rm Hz} \times N \times \left( \frac{m}{0.09\,\rm GeV} \right)^2 \ .
\end{eqnarray} 
The frequency window most sensitive to the LIGO experiment is 50--1000\,Hz. Therefore, if the $f_{\rm max}$ derived in Eqs.~(\ref{fmax1}) or (\ref{fmax2}) lies within this window and if the DSS pair is located close enough to the earth, LIGO has the potential to detect the gravitational waves. 
For the SU($N$) glueball model, the $m$-$N$ parameter space that can potentially be probed by the future running of LIGO is shown in Fig.~\ref{LIGO} by the red shaded band (for the large $N$ potential) and by the region between the dark-red dashed lines (for the $\phi^4$ potential). Here, we have set the value $\lambda=8\pi^2/(3N^2)$, which follows from comparing Eqs.~(\ref{V1}) and (\ref{VlargeN}). In this case, the two potentials make very similar predictions on the gravitational wave frequency.

Also shown in Fig.~\ref{LIGO} are the region of parameter space which allows the SU($N$) glueball to be a warm dark matter candidate (yellow band) or self-interacting dark matter (blue band), as discussed in detail in Ref.~\cite{Soni:2016gzf}. In particular, the dark matter self-interaction cross section is given by $\sigma_{2\to2}\sim\lambda^2/m^2$~\cite{Soni:2016gzf}, and here we assume the order one parameter $a_4$ in Eq.~(\ref{VlargeN}) to be in the range $1/3<a_4<3$.
On the other hand, the warm dark matter scenario can be achieved through the $3\to2$ annihilation process among the glueball particles (possible with the large $N$ potential) and the collisional damping~\cite{Soni:2016gzf}.
In the lower left corner of Fig.~\ref{LIGO}, the glueball dark matter has either too strong self-interaction or too large damping scale in the power spectrum, and the gray region is already ruled out by the bullet cluster and Lyman-$\alpha$ forest observations.

Interestingly, the current LIGO experiment is already probing the self interacting glueball dark matter with mass scale around 0.1\,GeV. The future gravitational wave observatories that are sensitive to lower frequencies (for example LISA/eLISA could probe 0.03\,mHz to 0.1\,Hz~\cite{lisa, Seoane:2013qna}) will be able to further probe the parameter space of a lighter (between keV and MeV) glueball dark matter, and even that of a warm dark matter. The region that could be potentially probed by eLISA is shown by the magenta band in Fig.~\ref{LIGO}.

\subsection{\large Conclusion and Outlook}

To summarize, in this work we explored a natural and important consequence of having glueball dark matter from a hidden sector with pure SU($N$) gauge symmetry --- the formation of dark SU($N$) stars (DSS). We solved the classical Einstein-Klein-Gordon equations for the mass-radius relations of the DSS. Because the dark glueball is a real scalar, the DSS configuration in general is time dependent and oscillates with a frequency given by the glueball mass. In our calculation, we take advantage of the large hierarchy where each DSS oscillation frequency is much higher than the gravitational wave frequency radiated by binary DSS (that can be observed, for example, by LIGO), and semi-analytically solve for the time-averaged DSS configuration.

Based on this calculation, we derive the frequency of gravitational waves radiated by binary DSS systems, as a function of the only two parameters in this simple model, the glueball mass $m$ and the number of colors $N$. We confront the model predictions to the frequency window sensitive to LIGO and future gravitational wave observatories, and find the regions of parameter space which could potentially be probed. This model offers an exciting connection between the gravitational wave radiation on compact dark stellar scales and the dark matter self-interaction on (dwarf-) galactic scales. Our main results are summarized in the key plot Fig.~\ref{LIGO}.

There are several further comments in order. 

\begin{itemize}

\item Throughout our discussions the scalar glueball dark matter is treated effectively as a scalar field. On the theoretical side, it is known to be not easy to generate a small mass for a fundamental scalar particle. A much more appealing way is to generate the mass dynamically in an $SU(N)$ gauge theory, where the dimensional transmutation is a well understood effect. On cosmological scales, at the moment we do not know how to differentiate our glueball from a fundamental scalar. This is an interesting and open issue to which we may return in the future.

\item We find the ratio of our glueball dark matter radius to its mass is $(R/M)_{\rm DSS} = {x_R}/({M_{pl}^2 \mathcal{M}_*(x_R)})$. In contrast, the ratio for a Schwarzschild black hole is simply $(R/M)_{\rm BH} = {2}/{M_{pl}^2}$. From our numerical calculation, we find that $x_R/\mathcal{M}_*(x_R)>2$ is always the case, thus the radius of the DSS is larger than the Schwarzschild radius of a black hole with equal mass. Therefore, a DSS will not collapse into a black hole. 

\item Because the mass-radius relation of our glueball dark star is different from that of a black hole, it may be possible to distinguish the binary glueball dark star from a binary black hole as the source for gravitational waves. The useful information in the gravitational wave spectrum for this precision measurement includes the time dependence of gravitational wave frequency, the amplitude, as well as the maximum frequency.  

\item One might also consider a binary system made of one black hole and one glueball dark star. The highest orbiting angular frequency (just before merging), $\omega$, satisfies $\omega^2 \simeq {2 G (M_{\rm BH} + M_{\rm DSS})}/{(R_{\rm BH} + R_{\rm DSS})^3}$.
Clearly, in the case when $M_{\rm BH} \gg M_{\rm DSS}$ and $R_{\rm BH} \gg R_{\rm DSS}$ (or the other way around), 
it is similar to the highest frequency for the case of binary black hole (binary DSS) merging.

\item Another potential way to distinguish the glueball dark stars and black holes is to study the gamma ray portion in the total energy loss (compared to the energy carried away by gravitational waves) during the merging process. This may be significant if the dark glueball is light thus its occupation number is high in the DSS, and if the dark glueball has strong enough coupling to the SM photon via higher dimensional operators. The gamma ray emissions can serve as the point-like source for locating the position of the DSS merger. We leave a more quantitative calculation of this interesting possibility to a future work.

\item Last but not the least, it is also an interesting and relevant question to investigate the formation and accretion process of the dark stars in the early universe and during galaxy formations, which will inform us the fraction of glueball dark matter in the form of the dark SU($N$) stars. The abundance of DSS depends on the primordial density perturbations in the glueball dark matter, and therefore is sensitive to the history of the very early universe. The relic abundance of the very massive DSS can impact the dynamics of (dwarf) galaxy formation and is strongly constrained~\cite{Brandt:2016aco}.

\end{itemize}

As an added note, a first-order phase transition of the dark $SU(N)$ sector in the early universe could also source gravitational waves~\cite{Schwaller:2015tja}. The corresponding frequency is typically much lower than that from binary dark stars considered in this work. 

\bigskip

\noindent{\it Acknowledgement.} We would like to thank John Terning and Enrico Rinaldi for discussions. 
The work of AS is supported in part by the DOE Grant No. DE-AC-02-98CH10886.
This work of YZ is supported by the DOE Grant No. DE-SC0010143. 
Y.Z. acknowledges the hospitality from the Aspen Center for Physics Aspen Center for Physics, which is supported by National Science Foundation grant PHY-1066293.

\end{document}